\documentclass[authoryear,preprint,review,12pt]{elsarticle}

\usepackage{graphicx}

\usepackage{amssymb}
\usepackage{amsthm}

\usepackage{lineno}
\linenumbers*[1]

\biboptions{authoryear,semicolon,sort,round}

\journal{Physica A}

\begin{document}

\begin{frontmatter}




 \title{First passage time statistics of Brownian motion with purely time dependent drift and diffusion}
 \author[pratt,nic]{A. Molini\corref{cor1}}
 \ead{annalisa.molini@duke.edu}
 \author[augsburg]{P. Talkner}
  \author[nic,pratt]{G.~G. Katul}
  \author[pratt,nic]{A. Porporato}

 \cortext[cor1]{Corresponding author}
 \address[pratt]{Department of Civil and Environmental Engineering, Pratt School of Engineering, Duke University, Durham, North Carolina, USA}
 \address[nic]{Nicholas School of the Environment, Duke University, Durham, North Carolina, USA}
 \address[augsburg]{Institut f\"{u}r Physik, Universit\"{a}t Augsburg, Augsburg, Germany}


%
%

\begin{abstract}
Systems where resource availability approaches a critical
threshold are common to many engineering and scientific
applications and often necessitate the estimation of first passage
time statistics of a Brownian motion (Bm) driven by time-dependent
drift and diffusion coefficients. Modeling such systems requires
solving the associated Fokker-Planck equation subject to an
absorbing barrier. Transitional probabilities are derived via the method of
images, whose applicability to time dependent problems is shown to
be limited to state-independent drift and diffusion coefficients
that only depend on time and are proportional to each other.  
First passage time statistics, such as the survival probabilities and
first passage time densities are obtained analytically. The
analysis includes the study of different functional forms of the
time dependent drift and diffusion, including power-law time
dependence and different periodic drivers. As a case study of
these theoretical results, a stochastic model of  
water resources availability in snowmelt dominated 
regions is presented, where both temperature effects and
snow-precipitation input are incorporated.

\end{abstract}


\end{frontmatter}



\section{Introduction}
\label{sec:introduction} A wide range of geophysical and
environmental processes occur under the influence of an external
time-dependent and random forcing. Climate-driven phenomena, such
as plant productivity \citep{ehleringer1997}, steno-thermal
populations dynamics \citep{mcclanahan2003}, crop production
\citep{rosenzweig1994}, the alternation between snow-storage and
melting in mountain regions \citep{marks1998,hamlet1999}, the life
cycle of tidal communities
\citep{barranguet1998,bertness1997,heather2009}, and water-borne
diseases outbreaks \citep{pascual2002,patz2005} offer a few such
examples. In particular, several environmental systems can be
described by state variables representing the
availability of a resource whose dynamics is forced by diverse environmental 
factors and climatic oscillations. 
Elevated regions water availability -- mainly originating from the melting of  snow masses
accumulated during the winter period (under the forcing of increasing temperatures), and
precipitation (moving from the solid precipitation to the rainfall regime) -- 
offers a relevant case study (presented in Section
\ref{sec:snowmelt}). All of these processes are now receiving
increased attention in several branches of ecology, climate sciences and hydrology, due to their
inherent sensitivity to climatic variability.

Analogous dynamical patterns can be found in slowly-driven,
non-equili-\\ brium systems with self organized criticality ({\it
SOC}), where the density of potentially relaxable sites in the
system can be described via a random walk with time-dependent
drift and diffusion terms \citep{adami1995, bak1995, jensen1998}.
In these systems, the time dependence in the diffusion term
derives from a gradual decrease of susceptible sites, so that
sites availability acts on the directionality and pathways (drift
term) of the ``avalanches''  till diffusion ``kills'' all the
activity in the system \citep[][pp. 120--131]{redner2001}. Similar dynamics occur in systems
displaying stochastic resonance, where noise becomes modulated by
an external periodic forcing \citep[see][and references
therein]{bulsara1994,bulsara1995,gammaitoni1998,macdonnell2008}.

In many instances, the above mentioned processes are restricted to
the positive semi-plane or to the time at which a certain critical
threshold is reached, and are represented by a Fokker-Planck (FP)
equation with an absorbing barrier. The main focus here is on the
first passage time statistics of the process, such as the survival
probabilities and the first passage time densities.
In the following, a brief review of the general properties of the
time-dependent drift and diffusion processes with an absorbing
barrier is presented. For constant drift and diffusion, the
conditional probabilities are usually obtained via the method of
images due to Lord Kelvin \citep[see][p. 340]{feller2}. The
applicability of this method to the solution of time-dependent
problems and its limitations are discussed and a necessary and 
sufficient criterion is formulated in
Section~\ref{subsec:absorbingbarrier}. The analysis is then
extended to different functional forms of the time-dependent drift
and diffusion terms. Section \ref{subsec:powerlaw} shows the
analytical results for the first passage time statistics for a
power-law time dependent drift and diffusion, while time-periodic
drivers are analyzed in Section~\ref{subsec:periodic}
\citep[see][and references therein, for a more comprehensive
review of periodically-driven stochastic
processes]{changho2010,jung1993,talkner2005}. Finally, in
Section~\ref{sec:snowmelt}, we present a stochastic model 
of the total mountain water equivalent during the apex phase of the melting season, 
incorporating both temperature effects and
snow-precipitation input in the form of a power-law time-dependent
Bm with an absorbing boundary.

\section{Modeling Framework}
\label{subsec:modelingframework}

When a time-dependent random forcing is the dominant driver of the
dynamics, a general representation for the state variable $x(t)$
can be formulated in the form of a stochastic differential
equation given by
\begin{equation}
dx(t) = \mu (t){\kern 1pt} \,dt + \sigma (t)\,dW(t)  \label{eq:ito}
\end{equation}
\noindent where $\mu (t)$ and  $\sigma (t)$ are purely
time-dependent drift and diffusion terms, and $W(t)$ is a Wiener
process with independent and identically Gaussian distributed (iid)
increments $W(t) - W(s) \sim \mathcal{N}(0,t - s)$ for all $t
\geqslant s \geqslant 0$. By assuming $t_0=0$, the solution
of~(\ref{eq:ito}) takes the form
\begin{equation}
x(t)  = x_0  + \int_{0}^t {\mu (s)ds + } \int_{0}^t {[\sigma (s)dW(s)]} \label{eq:itosol}
\end{equation}
where $t$ is time and $x_0=x(0)$ can be either a random or a
non-random initial condition independent of $W(t) - W(0)$. The
associated FP equation describing the evolution of the probability
density function (pdf) of $x(t)$ can be expressed as
\begin{equation}
\frac{{\partial p(x,t|x_0 )}}
{{\partial t}} =  - \mu (t)\frac{{\partial p(x,t|x_0 )}}
{{\partial x}} + \frac{1}
{2}\sigma ^2 (t)\frac{{\partial ^2 p(x,t|x_0 )}}
{{\partial x^2 }},
\label{eq:fpe}
\end{equation}
\noindent where $p(x,t|x_0 )$ is the transition pdf with initial
condition $\delta(x-x_0)$ at $t_0$. Eq.~(\ref{eq:fpe}) can
also be expressed as a continuity equation for probability
\begin{equation}
\frac{\partial }
{{\partial t}}p(x,t|x_0 ) = -\frac{\partial }
{{\partial x}}j(x,t|x_0 ),
\label{eq:fpecurrent}
\end{equation}
where
\begin{equation}
 j(x,t|x_0 ) =  \mu (t)\,p(x,t|x_0 ) - \frac{1}{2}\sigma ^2 (t)\frac{{\partial p(x,t|x_0 )}}{{\partial x}} ,
 \label{eq:current1}
\end{equation}
 \noindent is the probability current (or flux) and $p(x,t|x_0 )$ is the conditional probability.
The solution of the FP equation in (\ref{eq:fpe}), is usually approached numerically (see for e.g., \citet{schindler2005}).
Whether this equation is analytically solvable for different functional forms
of $\mu (t)$ and $\sigma (t)$ with an absorbing boundary, and
whether these solutions can be applied in the study of the first
passage statistics at such boundary is the main focus of this
work. Case studies that employ these solutions are also presented.




\subsection{Solution with Natural Boundaries}
\label{subsec:naturalboundaries} Consider first the solution of
the FP equation (\ref{eq:fpe}) in the unbounded case.
Given that the drift and diffusion coefficients depend only on time, the
parabolic equation~(\ref{eq:fpe}) can still be reduced to a
constant-coefficient equation of the form
\begin{equation}
\frac{{\partial p(z,\tau )}}
{{\partial \tau }} = \frac{{\partial ^2 p(z,\tau )}}
{{\partial z^2 }}
 \label{eq:pdetransformed}
\end{equation}
by transforming the original variables $x$ and $t$ into
\begin{equation}
\tau  = \frac{1}
{2}\int {\sigma ^2 (t)dt + A}
 \label{eq:tau}
\end{equation}
\noindent and
\begin{equation}
z = x - \int {\mu (t)dt + B}
 \label{eq:zeta}
\end{equation}
where $A$ and $B$ are generic constants. The solution with natural boundaries is then
\citep{polyanin2002} $p(z,\tau ) = \frac{1}{{2\sqrt {\pi \tau }
}}\exp \left( { - \frac{{z^2 }}{{4\tau }}} \right)$. Hence, given
the initial condition 
\begin{equation}
p(x,0|x_0) = \delta (x - {x_0}), \label{eq:dirac}
\end{equation}
\noindent the following normalized solution for an unrestricted process, starting from $x_0$, can be obtained as
\begin{equation}
p_u(x,t|x_0) = \frac{1}
{{2\sqrt {\pi S(t)} }}\exp \left[- \frac{{(x - x_0  - M(t))^2 }} {{4S(t)}} \right],
 \label{eq:solunbounded}
\end{equation}
where, assuming the integrability of $\mu (t)$ and $\sigma (t)$,
\begin{equation}
M(t) = \int_0^t {\mu (s)ds}
 \label{eq:emme}
\end{equation}
and
\begin{equation}
S(t) = \frac{1}
{2}\int_0^t {\sigma ^2 (s)ds }.
 \label{eq:esse}
\end{equation}
It should be noted that the transformation in equations ~(\ref{eq:zeta}) and~(\ref{eq:tau}) 
also applies to any boundary condition imposed at a finite position.
Therefore, as will be seen, it is not directly helpful in solving first passage time problems, 
as in that case it  would lead to a problem with {\it moving} absorbing boundary conditions.


\subsection{First Passage Time Distributions}
\label{subsec:absorbingbarrier} For a Bm process commencing at a
generic position $x_0$ at $t=0$, the time at which this process
reaches an arbitrary threshold $a$ for the first time (first
passage time) is itself a random variable whose statistics are
fundamental in many branches of science such as chemistry,
neural-sciences and econometrics. In the following, it is assumed
that the process is starting at a certain state $x_0>0$ and that
it is bounded to the positive semi-axis via an absorbing barrier
$x=0$. This hypothesis does not imply any loss of generality,
considering that the solution of Eq.~(\ref{eq:fpe}) with an
absorbing boundary condition only depends on the distance of the initial point $x_0$
from the threshold, but not separately on $x_0$ and the threshold position.
 Eq. (\ref{eq:fpe}) is then solved with the boundary
condition
\begin{equation}
p(0 ,t) = 0 ,
 \label{eq:absorbingbarriercondition}
\end{equation}
\noindent and the additional condition of $x=+ \infty $ being a
natural boundary to ensure that $ j( + \infty ,t\left| {x_0 )}
\right.$ = $0$. For such a system, the survival probability
$F(t\left| {x_0 )} \right.$ is defined as the probability of the
process trajectories not absorbed before time $t$, i.e.
\begin{equation}
F(t\left| {x_0 )} \right. = \int_0^{ + \infty } {p (x,t\left| {x_0 ){\kern 1pt} \,dx} \right.}
\label{eq:Fsurvival}
\end{equation}
\noindent and the first passage probability density $g(t\left|
{x_0 )} \right.$ is either the ``rate of decrease'' in time of $F$
\begin{equation}
g(t\left| {x_0 )} \right. =  - \frac{\partial }
{{\partial t}}F(t\left| {x_0 ){\kern 1pt} \,} \right.
\label{eq:g1}
\end{equation}
\noindent or, alternatively, the negative probability current at the
boundary
\begin{equation}
g(t\left| {x_0 } \right.)= \frac{\sigma^2(t)}{2} \frac{\partial }{{\partial x}} p(x,t\left| {x_0} \right.)\left| {_{x = 0} } \right. ,
    \label{eq:g2}
\end{equation}
\noindent since $p(0,t|x_0)=0$ from~(\ref{eq:absorbingbarriercondition}).

\subsection{Method of Images in Time-Dependent Systems}
\label{subsec:imagemethod} When the drift and diffusion terms are
independent of $t$ and $x$, Eq. (\ref{eq:fpe}) with absorbing
boundaries can be readily solved by the method of images, often
adopted in problems of heat conduction and diffusion
\citep{cox1965, daniels1982, redner2001,lo2002}. This method can
also be used for solving boundary-value problems for a Bm with
particular forms of time-dependent drift and diffusion.  The basic
premise of this method is that given a linear PDE with a point source (or
sink) subject to homogeneous boundary conditions in a finite
domain, its general solution can be obtained as a superposition
of many `free space' solutions (i.e. disregarding the boundary
conditions) for a number of virtual sources (i.e. outside the
domain) selected so as to obtain the correct boundary
condition.  The image source (or sink) is placed as mirror image of the original source (or sink)
from the boundary with a strength or intensity selected to match
the boundary condition.

Consider equation (\ref{eq:fpe}) with the conditions (\ref
{eq:dirac}) and  (\ref{eq:absorbingbarriercondition}). 
To solve this problem with the method of images, the barrier at 0 is replaced by a mirror source located at a generic point $x=y$, with $y<0$ such that the solutions of the Fokker-Planck equation emanating from the original and mirror sources exactly compensate each other at the position of the barrier at each instant of time \citep{redner2001}. 
This implies the initial conditions in
(\ref {eq:dirac}) must now be modified to
\begin{equation}
p(x,0) = \delta (x - x_0 ) - \exp \left(- \eta\right) \delta (x -y ),
 \label{eq:initialmod}
\end{equation}
\noindent where $\eta$ determines the strength of the mirror image source. Due to the
linearity of the FP equation, the solution in the presence of the
initial condition (\ref{eq:initialmod}) is the superposition of
elementary solutions
\begin{equation}
p(x,t|x_0 ) = p_u (x,t|x_0 ) - \exp \left(- \eta\right) p_u(x,t|y).
 \label{eq:superimposition}
\end{equation}
\noindent Since the condition (\ref{eq:absorbingbarriercondition}) requires that $p(0,t|x_0 ) =0$, one obtains that
\begin{equation}
\frac{{(M(t) + x_0 )^2 }}
{{4S(t)}} = \eta + \frac{{(M(t) + y)^2 }}
{{4S(t)}}
 \label{eq:compareexp}
\end{equation}
\noindent for all $t>0$. By assuming $t=0$, we have $x_0^2=y^2$ and recalling that $y<0$,
the resulting image position is $-x_0$. This, inserted again in
Eq. (\ref{eq:compareexp}), yields
\begin{equation}
\frac{{M(t)}}
{{S(t)}} = \frac{\eta }
{{x_0 }} = q ,
 \label{eq:peclet}
\end{equation}
\noindent where the constant $q$ is analogous to the P{\'e}clet
number of the process -- i.e. the ratio between the advection and
diffusion rates \citep{redner2001}.

After differentiating (\ref{eq:peclet}) with respect to $t$, it is seen that the
method of images requires that the drift and the diffusion terms
be proportional to each other.
Namely, the intensity $\eta$ of the image source must be constant
in time. In fact, only in this case it is still possible to
transform the original time scale into a new one, for which the
transformed process is governed by time-independent drift and
diffusion terms. Hence, writing the drift and diffusion terms as
\begin{equation}
\mu (t) = kh(t)\;\;\;\;{\rm{and}}\;\;\;\;\frac{{\sigma ^2 }}
{2} = lh(t),
 \label{eq:musigmanew}
\end{equation}
\noindent the associated FP equation is
\begin{equation}
\frac{{\partial p}}
{{\partial t}} = h(t)\left( { - k\frac{\partial }
{{\partial x}} + l\frac{{\partial ^2 }}
{{\partial x^2 }}} \right)p.
 \label{eq:fpe2}
\end{equation}
\noindent Transforming the original time $t$ variable in
\begin{equation}
\tilde \tau  = \int_0^t {h(s)ds}
 \label{eq:timetransformation2}
\end{equation}
\noindent Equation~(\ref{eq:fpe2}) finally becomes
\begin{equation}
\frac{{\partial p}}
{{\partial \tilde \tau }} = \left( { - k\frac{\partial }
{{\partial x}} + l\frac{{\partial ^2 }}
{{\partial x^2 }}} \right)p.
 \label{eq:fpe3}
\end{equation}
\noindent This condition is valid for any time-dependent diffusion when the
drift is identically vanishing.
Assuming the proportionality in (\ref{eq:peclet})
between $\mu(t)$ and $\sigma(t)$, the general solution for
(\ref{eq:fpe}) under conditions (\ref {eq:dirac}) and
(\ref{eq:absorbingbarriercondition}) can be written as
\begin{equation}
\begin{array}{ll}
  p(x,t|x_0 ) &= \frac{1}
{{2\sqrt {\pi S(t)} }}\left\{ {\exp \left[ - \frac{{(x - x_0  - M(t))^2 }}
{{4S(t)}} \right] } \right. \hfill \\
 & \left. { - \exp \left( - {x_0 q} \right) \ \exp \left[ - \frac{{(x + x_0  - M(t))^2 }}
{{4S(t)}}\right] } \right\}, \hfill \\
\end{array}
 \label{eq:generalabsorbing}
\end{equation}
\noindent provided $M(t)=q S(t)$.
Substituting for constant drift and diffusion in
(\ref{eq:generalabsorbing}) one recovers the well-known solution
for a biased Bm \citep{cox1965}
\begin{equation}
\begin{array}{ll}
 p(x,t|{x_0}) = &\frac{1}{{\sqrt {2\pi } \sqrt {{\sigma ^2}} t}}\left\{ {{\exp \left[ - \frac{{{{({x_0} - x + \mu t)}^2}}}{{2{\sigma ^2}t}}\right]}  } \right. \\
 & \left. { - {\exp \left( - \frac{{2{x_0}\mu }}{{{\sigma ^2}}}\right)}\ {\exp \left[ \frac{{{{(x + {x_0} - \mu t)}^2}}}{{2{\sigma ^2}t}}\right] }}\right\} \\
 \end{array}
 \label{eq:pbiasedBM}
\end{equation}
with survival function $F(t|{x_0})$ given by
\begin{equation}
F(t|{x_0})  = \Phi \left\{ {\frac{{\mu t + {x_0}}}{{\sigma \sqrt t }}} \right\} - \exp \left(- \frac{{2{x_0}\mu }}{{{\sigma ^2}}}\right) \Phi \left\{ {\frac{{\mu t - {x_0}}}{{\sigma \sqrt t }}} \right\} ,
 \label{eq:SbiasedBM}
 \end{equation}
 \noindent where $\Phi$ is the standard normal integral, and first passage time distribution
\begin{equation}
g(t|{x_0}) = \frac{{{x_0}}}{{\sigma \sqrt {2\pi } {t^{3/2}}}}\ {\exp \left[ - \frac{{{{({x_0} + \mu t)}^2}}}{{2{\sigma ^2}t}}\right]}.
\label{eq:gbiasedBM}
\end{equation}
Equation~(\ref{eq:gbiasedBM}) is the Wald (or inverse Gaussian)
density function, that for a zero drift becomes of order
$t^{-3/2}$ as $t \to  + \infty $ (the first passage time has no
finite moments for pure diffusion). 


Similarly, the solution to the FP in equation (3) with a reflecting boundary at $x=0$ 
can be obtained by the method of images provided that drift and diffusion are proportional to each other. 
The solution then becomes 
\begin{equation}
\begin{array}{ll}
  p(x,t|x_0 ) &= \frac{1}
{{2\sqrt {\pi S(t)} }}\left\{ {\exp \left[ - \frac{{(x - x_0  - M(t))^2 }}
{{4S(t)}}\right] + \exp \left( - {x_0 q}\right)\exp \left[ - \frac{{(x + x_0  - M(t))^2 }}
{{4S(t)}}\right] } \right. \hfill \\
 & \left. { -\frac{1}
{2}\frac{{M(t)}}
{{S(t)}}\exp \left(\frac{{x\eta}}
{{x_0}}\right) \left[ {1 - {\rm{erf}}\left( {\frac{{x + x_0  + M(t)}}
{{2\sqrt {S(t)} }}} \right)} \right]} \right\}  , \hfill \\
\end{array}
 \label{eq:generalreflecting}
\end{equation}
\noindent with $ \frac{1}{2}\left[ {1 - {\rm{erf}}\left( {\frac{{x
+ x_0  + M(t)}}{{2\sqrt {S(t)} }}} \right)} \right]$ being the
Q-function representing the tail probability of a Gaussian
distribution. Equation~(\ref{eq:generalreflecting}) generalizes the solution in \citet{cox1965} for a Bm
with constant drift and diffusion and a reflecting boundary at
$0$.

\section{Time Dependent Drift and Diffusion}
\label{sec:timedependent}
\subsection{Power-Law Time Dependence}
\label{subsec:powerlaw} 
\begin{figure}
\noindent\includegraphics[width=35pc]{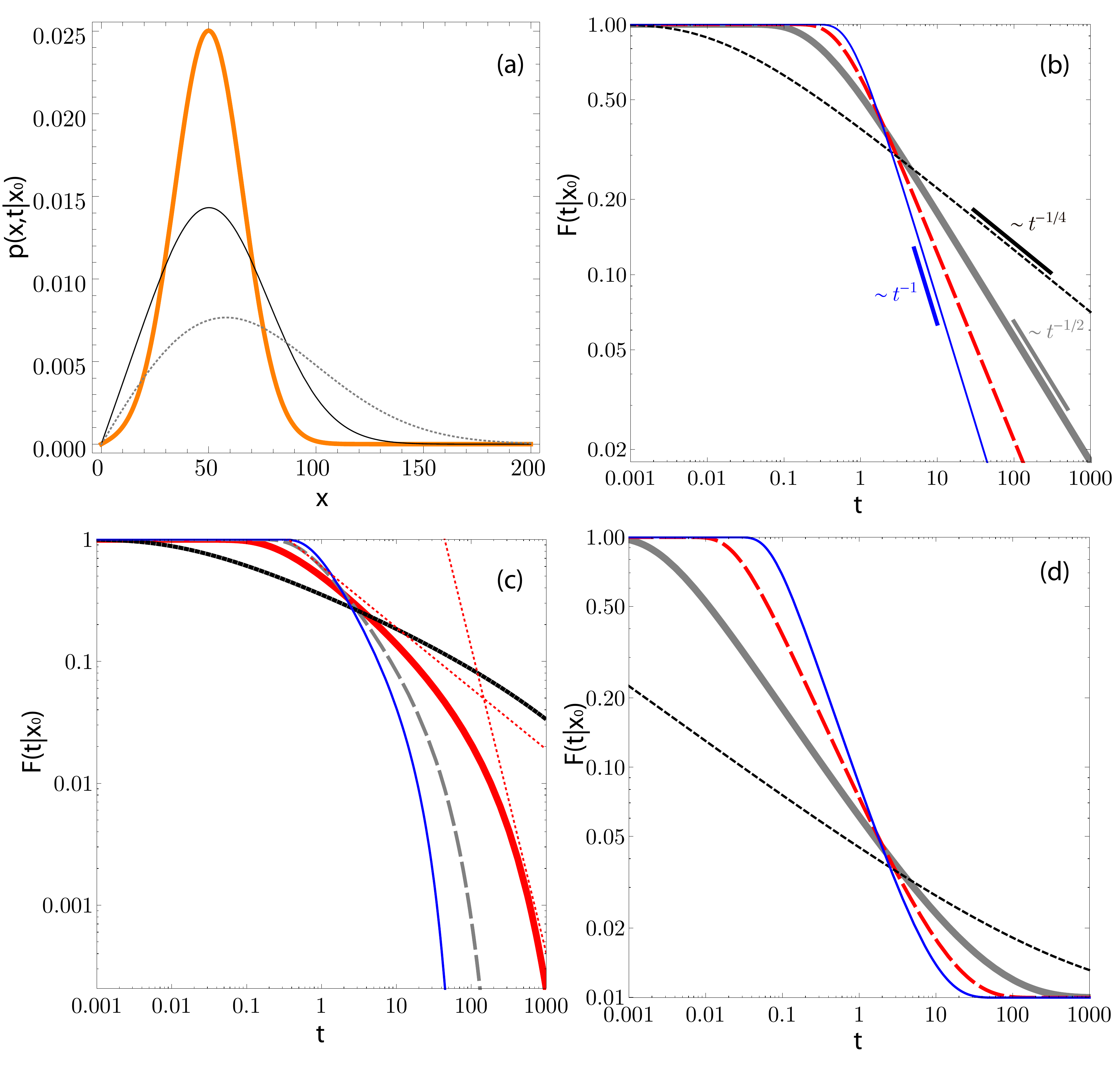}
\caption{Conditional probability $p(x,t|x_0)$ at different fixed times $t$ (a) and survival function $F(t|x_0)$ (b) for the pure power-law time dependent process described in Section \ref{subsec:powerlaw}, together with $F(t|x_0)$ for the negatively biased power-law process (c) and for the positively biased one (d). Panel (a) represents $p(x,t|x_0)$ at a fixed time $t=15$ steps for $A=15$, $x_0=50$,  and $\alpha$  $=-0.1$ (bold line), $=0.5$ (thin line), and $=1$ (dotted line). In (b) $F(t|x_0)$ is displayed as a function of $t$ for $A=1$, $x_0=1$ and $\alpha=0$ (constant diffusion, bold line), $\alpha=-0.5$ (thin dotted line),  $\alpha=0.5$ (dashed line), and  $\alpha=1 $ (thin line). Panels (c) and (d) display respectively a negatively biased power-law time dependent Bm and a positively biased one for the same set of parameters in (b) and $q=-0.1$ and $q=0.1$.}
 \label{Fpowerlaw}
\end{figure}

As a first example of Bm with purely
time-dependent drivers, the case of an unbiased diffusion ($q=0$)
and power-law time dependent diffusion term $\sigma^2(t)  = 2
A{t^{\alpha}}$ and $\alpha>-1$ are considered. For this
process, the conditional probability $p(x,t|x_0)$ with absorbing
barriers at $0$, takes on the form
\begin{equation}
\begin{array}{ll}
 p(x,t\left| {{x_0}} \right.) =& 
 \frac{{\sqrt {1 + \alpha } {\mkern 1mu} }}{{2\sqrt {A\pi } }}{t^{ - (\alpha  + 1)}}\left\{ {{\exp \left[ - \frac{{{t^{ - (\alpha  + 1)}}{{(x - {x_0})}^2}(1 + \alpha )}}{{4A}}\right]}} \right.  \\
 &\left. { - {\exp \left[ - \frac{{{t^{ - (\alpha  + 1)}}{{(x + {x_0})}^2}(1 + \alpha )}}{{4A}}\right]}} \right\} ,\\
 \end{array}
 \label{eq:ppowerlawpurediff}
\end{equation}
\noindent while the survival function becomes
\begin{equation}
F(t|{x_0}) = {\rm{erf}}\left( {\frac{{{x_0}{{(1 + \alpha )}^{\frac{1}{2}}}}}{{2\sqrt A }}{t^{ - \frac{{\alpha  + 1}}{2}}}} \right).
 \label{eq:Spowerlawpurediff}
\end{equation}
Figure~\ref{Fpowerlaw}(a) shows the conditional probability
(\ref{eq:ppowerlawpurediff}) at a fixed time instance $t=15$ time
steps for $A=15$, $x_0=50$,  and $\alpha=$  $-0.1$ (bold line)
$0.5$ (thin line), and $1$ (dotted line). Given the asymptotic
properties of the error function \citep{abramowitz1964}, the
long-time behavior of  $F(t|x_0)$ is then $\sim \frac{{{x_0}{{(1 +
\alpha )}^{{\raise0.5ex\hbox{$\scriptstyle 1$}
\kern-0.1em/\kern-0.15em \lower0.25ex\hbox{$\scriptstyle
2$}}}}}}{{2\sqrt A }}{t^{ - \frac{{\alpha  + 1}}{2}}}$, recovering
for $\alpha=0$ the $-1/2$ tail decay of an unbiased constant
diffusion (see Figure~\ref{Fpowerlaw}(b)). Also, by
differentiating Eq.~(\ref{eq:Spowerlawpurediff}), one obtains
\begin{equation}
g(t|{x_0}) = \frac{{{x_0}}}{{2\sqrt {\frac{{A\pi }}{{{{(\alpha  + 1)}^3}}}} {t^{(3+\alpha)/2}}}}{\exp \left[- \frac{{{x_0}(\alpha  + 1){\kern 1pt} {t^{ - (\alpha  + 1)}}}}{{4A}}\right] }
 \label{eq:gpowerlawpurediff}
\end{equation}
\noindent whose tail behaves as $\sim t^{ - \left( {\frac{{3 +
\alpha }}{2}} \right)} $. Hence, Eq.~(\ref{eq:gpowerlawpurediff}) is an
inverse Gaussian distribution -- that for $\alpha =0$ becomes an inverse Gamma distribution with 
shape parameter $1/2$ \citep[][pp. 284--285]{johnson1994}.
These solutions characterize inter-arrival times between
intermittent events when a system displays sporadic randomness
\citep{gaspard1988, molini2009,rigby2010}. 

The solutions in the
case of proportional power-law diffusion and drift can be derived
in an analogous manner. For  $\mu(t)  = qA{t^\alpha }$ and
$\sigma(t) = \sqrt 2 {A^{1/2}}{t^{\alpha /2}}$, the
conditional probability $p(x,t|x_0)$ takes the form
\begin{equation}
\begin{array}{ll}
 p(x,t\left| {{x_0}} \right.) = &\\
  \frac{{\sqrt {\alpha  + 1} {\mkern 1mu} {t^{ - (\alpha  + 1)/2}}}}{{2\sqrt {A\pi } }}\left\{{\exp \left[ - \frac{{(1 + \alpha ){t^{ - (\alpha  + 1)}}{{\left( { - x + \frac{{Aq{t^{1 + \alpha }}}}{{1 + \alpha }} + {x_0}} \right)}^2}}}{{4A}}\right]} \right. &\\
\left.  - {\exp \left[ - q{x_0} - \frac{{(1 + \alpha ){t^{ - (\alpha  + 1)}}{{\left( {x - \frac{{Aq{t^{1 + \alpha }}}}{{1 + \alpha }} + {x_0}} \right)}^2}}}{{4A}}\right]}\right\} &\\
 \end{array} \label{eq:ppowerlaw}
\end{equation}
\noindent and the survival function, now incorporating the drift contribution, can be written as
\begin{equation}
\begin{array}{ll}
 F(t|{x_0}) = & \Phi \left\{ {{t^{ - \frac{{(\alpha  + 1)}}{2}}}\frac{{\left( {Aq{t^{\alpha  + 1}} + {x_0} + {x_0}\alpha } \right)}}{{2\sqrt {A(\alpha  + 1)} }}} \right\}   \\
&  - {\exp \left( - q{x_0}\right)\ }\Phi \left\{ {{t^{ - \frac{{(\alpha  + 1)}}{2}}}\frac{{\left( {Aq{t^{\alpha  + 1}} - {x_0} - {x_0}\alpha } \right)}}{{2\sqrt {A(\alpha  + 1)} }}} \right\}. \\
 \end{array}
\label{eq:Spowerlaw}
\end{equation}
For positive $q$'s,  $F(t|{x_0})$  tends in the long term to
$1-\exp (-qx_0)$, while for negative $q$'s, $F(x,t|x_0 ) \sim
\frac{{2\sqrt {\alpha  + 1} }} {{q\sqrt A }}t^{ - \frac{{\alpha  +
1}}{2}} \exp \left( - \frac{{q\sqrt {A\,} t^{\frac{{\alpha  + 1}}{2}}
}}{{2\sqrt {\alpha  + 1} }}\right)$. This fact implies that the
probability for a trajectory to be eventually absorbed is 1 for the biased process
directed towards the barrier, and $\exp (-qx_0)$ when the bias is
away from the barrier (infinite aging).
When the state variable represents the availability of a resource in time, the sign of $q$ determines if this resource is subject to continuos accumulation
(positive $q$), or it undergoes a total depletion (negative $q$) with probability $1$.
Such a result is analogous
to the one of a simple biased Bm with constant drift and diffusion
\citep{redner2001}, with the difference that in this case, $F(t|{x_0})$ decays to
0 or $1-\exp (-qx_0)$ with a rate that is governed by $\alpha$.

As an example, Figures \ref{Fpowerlaw} (c) and (d) respectively
show a negatively biased power-law time-dependent Bm and a
positively biased one for the same set of parameters in (b) and
$q=-0.1$ and $q=0.1$, for $A=1$, $x_0=1$ and $\alpha=0$ (constant
diffusion, bold line), $-0.5$ (thin dotted line),  $0.5$ (dashed
line), and $1$ (thin line). As evident in panel (c), $F(x,t|x_0 )$
presents a faster decay to zero with increasing $\alpha$, while
for the positively biased Bm in panel (d) the decay to the
asymptotic value $1-\exp (-qx_0)$ is slower with decreasing $\alpha$.

Finally, $g(t\left|x_0\right.)$ can be obtained from (\ref{eq:Spowerlaw}) as
\begin{equation}
g(t\left|x_0\right.)=\frac{x_0(1+\alpha )^{3/2}}{2\sqrt{\pi A}t^{\frac{3+\alpha}{2}}} \exp \left[-\frac{t^{-(\alpha +1)}\left(Aqt^{\alpha +1}+x_0+\alpha x_0\right){}^2}{4A(1+\alpha )}\right]
\label{eq:gpowerlaw}
\end{equation}
where for $\alpha=0$ the decay of $g(t|x_0)$ recovers the constant diffusion $t^{-3/2}$-law for $t \to \infty$ and $q=0$. 

\subsection{Periodic Drift and Diffusion}
\label{subsec:periodic} 
\begin{figure}
\noindent\includegraphics[width=30pc]{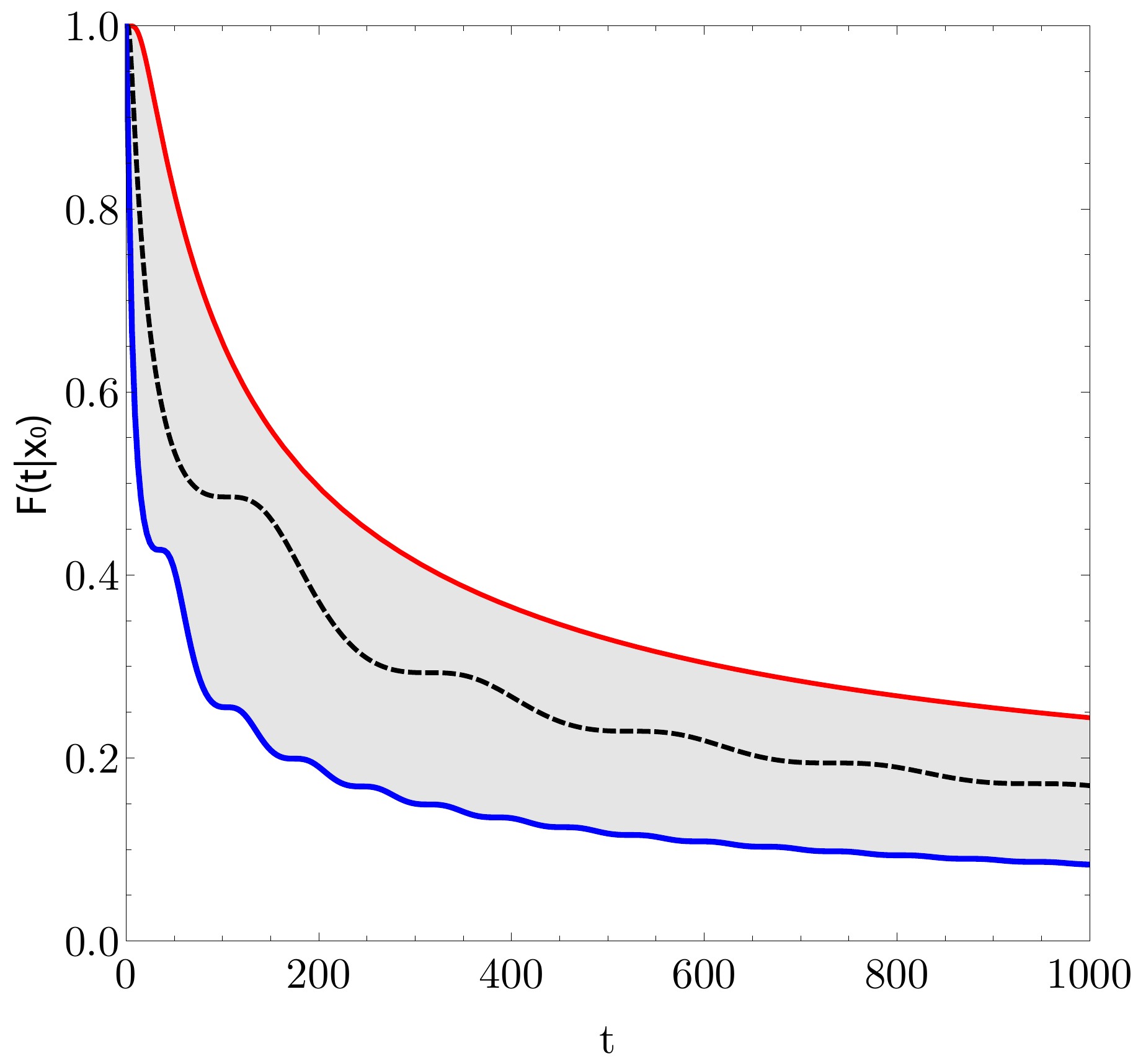}
\caption{Survival function $F(t|x_0)$ for the periodic purely diffusive process described in Section \ref{subsec:periodic} , and for $A=15$ and $x_0=50$.
Upper, dashed and lower curves represent $F$ for $\omega=0.0001$, $\omega=0.015$, and $\omega=0.045$, respectively.}
 \label{FFtpurediffusionperiodic}
\end{figure}

In this section, the case of a periodic
diffusion in the form $\sigma^2 (t)= \left[ {2A\cos (\omega t)}
\right]^2 $ and $q=0$ is considered. For periodically driven
diffusion, the conditional probability can be derived
in the form
\begin{equation}
\begin{array}{ll}
p(x,t\left| {x_0 } \right.) =& \left( {\frac{\omega }
{{\pi \vartheta(t) }}} \right)^{\frac{1}
{2}} \exp \left[ - \frac{{2\omega \left( {x^2  + x_0 } \right)}}
{\vartheta(t) }\right]\ \left\{ { \exp \left[\frac{{\omega (x + x_0 )^2 }}
{\vartheta(t) } \right] } \right. \\
&\left. {- \exp \left[\frac{{\omega (x - x_0 )^2 }}
{\vartheta(t) }\right] } \right\} \\
\label{eq:pdiffperiodic1}
\end{array}
\end{equation}
\noindent where $ \vartheta(t)  = A^2 [2\omega t + \sin (2\omega t)]
> 0 $.  Thus, the solution becomes modulated in time with frequency
$\omega$. The survival probability is in turn
\begin{equation}
F(t|x_0 )  = {\rm{erf} }\left( {x_0 \sqrt {\frac{\omega }{\vartheta(t) }} } \right) ,
\label{eq:Sdiffperiodic1}
\end{equation}
\noindent that is represented in
Figure~\ref{FFtpurediffusionperiodic} for different values of the
frequency $\omega$. Finally, the first passage time density
is an $\omega$-modulated inverse Gaussian distribution
\begin{equation}
g(t|x_0 ) = \frac{{4x_0 A^2 }}
{{\sqrt \pi  }}\frac{{\omega ^{3/2} {\rm{cos}}(\omega t)^2 }}
{{\vartheta(t) ^{3/2} }} \exp \left(- \frac{{\omega x_0^2 }}
{\vartheta(t) }\right).
\label{eq:gdiffperiodic1}
\end{equation}

\begin{figure}
\noindent\includegraphics[width=35pc]{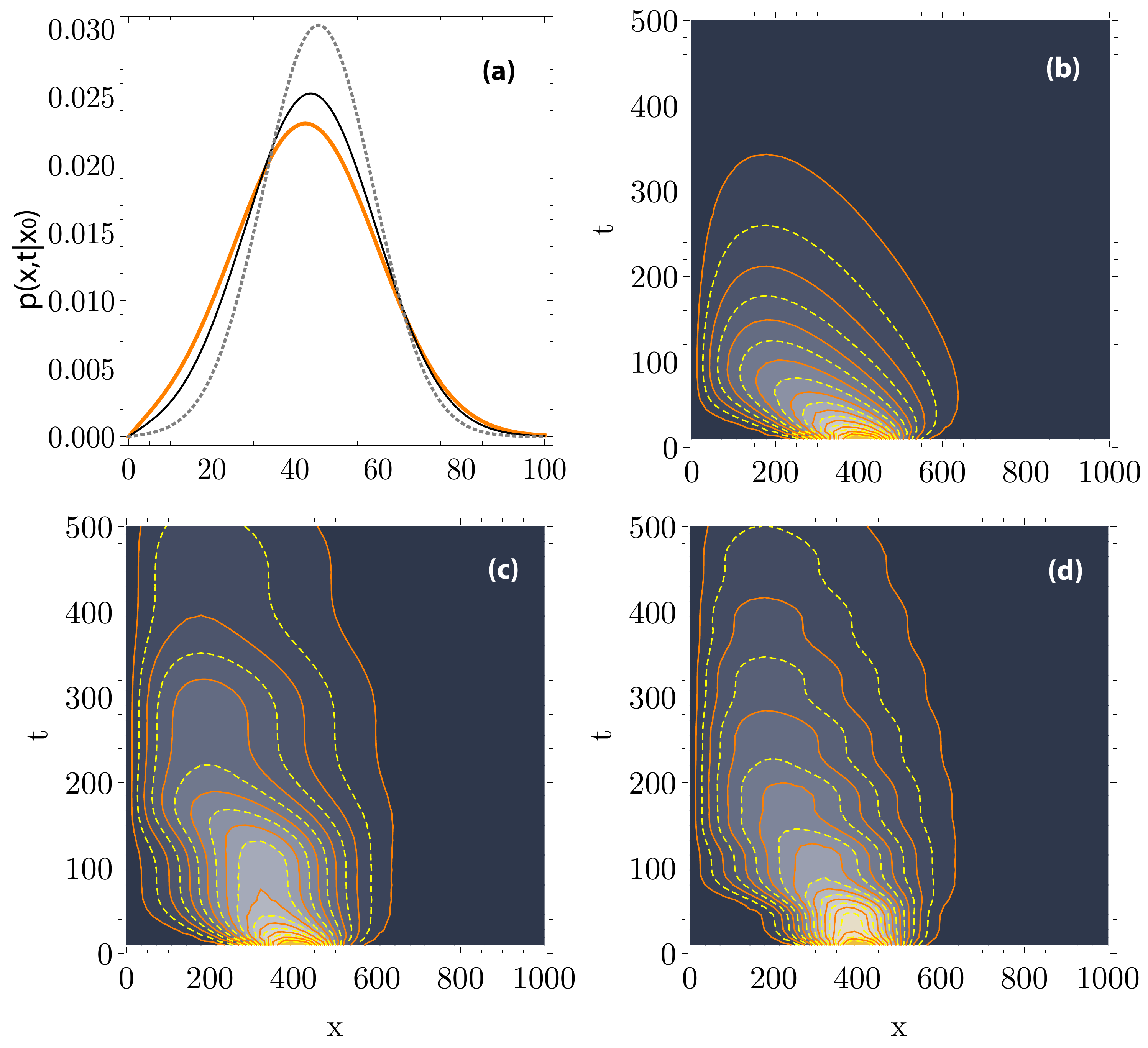}
\caption{Conditional probability $p(x,t|x_0)$ for the periodic negatively biased Bm described in Section \ref{subsec:periodic}. Panel (a) represents $p(x,t|x_0)$ at a fixed time $t=3$ steps for $A=15$, $x_0=50$, $q=-0.05$, and $\omega$  $=0.0001$ (bold line), $=0.5$ (thin line), and $=0.9$ (dotted line). Also, contour plots (b) to (d) show $p(x,t | x_0)$ for $A=15$, $x_0=450$ and $q=-0.01$ as a function of $x$ and $t$, for $\omega=0.0001$ (panel (b)), $\omega=0.015$ (panel (c)), and $\omega=0.045$ (panel (d)) respectively. Note how the negative drift forces the probability mass toward the barrier.}
 \label{Fperiodicdriftnegative}
\end{figure}

In the case $q\neq0$, the conditional probability $p(x,t\left| {x_0 } \right.)$ becomes
\begin{equation}
\begin{array}{ll}
p(x,t\left| {x_0 } \right.) = &\frac{{\sqrt \omega  }}
{{\sqrt {\pi q\vartheta(t) } }}\left\{ {\exp \left[ - \frac{{\omega \left( {x_0  - x + \frac{{q\vartheta(t) }}
{{4\omega }}} \right)^2 }}
{\vartheta(t) }\right] } \right. \\
&\left. { - \exp \left[ - qx_0  - \frac{{\omega \left( {x_0  + x - \frac{{q\vartheta(t) }}
{{4\omega }}} \right)^2 }}
{\vartheta(t) }\right] } \right\} ,\\
\end{array}
\label{eq:pperiodic1}
\end{equation}

\noindent where, again, the absorption at the barrier represents a
recurrent ($q<0$) or a
transient ($q>0$) state, as
was observed for the power-law drift and diffusion process in
Section~\ref{subsec:powerlaw}. 
The recurrent case is illustrated in
Figure~\ref{Fperiodicdriftnegative} (b)-(d), where we report the
time-position evolution of $p(x,t|x_0)$ as a function of
increasing $\omega$.
From~(\ref{eq:pperiodic1}), given
$\frac{\omega }{\theta} > 0$, the expression for the survival
function can be derived and takes the form
\begin{equation}
\begin{array}{ll}
  F(t|x_0 ) & = \frac{1}
{2}\left[ {1 + {\rm{erf}}\left( {\frac{{q\vartheta(t)  + 4x_0 \omega }}
{{4\sqrt {\omega \vartheta(t) } }}} \right)} \right. \hfill \\
  &\left. { + \exp\left(- qx_0 \right){\rm{erfc}}\left( {\frac{{q\vartheta(t)  - 4x_0 \omega }}
{{4\sqrt {\omega \vartheta(t) } }}} \right) - 2\exp\left(- qx_0 \right) } \right], \hfill \\
\end{array}
\label{eq:Speriodic1}
\end{equation}
\noindent which, given the equality ${\rm{erfc}}( - x) = 2 - {\rm{erfc}}(x)$, can be alternatively expressed as
\begin{equation}
F(t|x_0 ) = \Phi \left\{ {\frac{{q\vartheta(t)  + 4x_0 \omega }}
{{2\sqrt {2\omega \vartheta(t) } }}} \right\} - \exp\left(- qx_0 \right) \Phi \left\{ {\frac{{q\vartheta(t)  - 4x_0 \omega }}
{{2\sqrt {2\omega \vartheta(t) } }}} \right\} .
\label{eq:Speriodic33}
\end{equation}
\noindent The first passage time density  $g(t|x_0)$ is given by
\begin{equation}
g(t\left| {x_0 } \right.) = \frac{{4A^2 x_0 }}
{{\sqrt \pi  }}\frac{{\omega ^{^{3/2} } \cos (\omega t)^2 }}
{{\vartheta(t) ^{^{3/2} } }}\exp\left[ - \frac{{\left( {q\vartheta(t)  + 4x_0 \omega } \right)^2 }}
{{16\omega \vartheta(t) }} \right].
\label{eq:gperiodic1}
\end{equation}

The method of images can also be applied to the solution of
different forms of periodic drivers, such as the case $\mu(t) =
q(B + A{\rm{cos}}(\omega t))$ and $\sigma(t)  = \sqrt {2(B +
A{\rm{cos}}(\omega t))} $, with $(B + A{\rm{cos}}(\omega t))>0$.
In this last case, the drift term is the same as the one usually
investigated in neuron dynamics by simple integrate-and-fire
models displaying stochastic resonance \citep[see for example the
neuron dynamics case in][]{bulsara1994,bulsara1995}. In those
models, the diffusion is usually constant so that the condition in
equation (20) is not satisfied. Thus, it is often implied that
$\mu(t)<<\sigma^2/2$ to approximately resemble a time dependent
diffusion with drift identically vanishing or that $B>>A$ 
(approximating the simpler constant drift and diffusion case). In these cases, the
method of images only offers approximated solutions
\citep{bulsara1994,bulsara1995}). Specifically, for a time dependent (and
periodic) drift $\mu(t)=B  + A\cos (\omega t)$ and constant
diffusion $\frac{1}{2}\sigma^2$, an approximation for \(p
(x,t\left| {x_0 } \right. ) \) in the presence of an absorbing
barrier at \(0\) can still be obtained by using the method of
images conditional to the fact that $\mu(t)<<\sigma^2/2$. 
Only by adopting this assumption in fact, we can obtain an (approximated) solution for the survival function 
by means of Eq.~\ref{eq:generalabsorbing} although drift and diffusion are not strictly proportional to each other. In this way we find 
\begin{equation}
\begin{array}{ll}
  F(t|x_0 ) = &\frac{1}
{2}\left\{ {{\rm{erfc}}\left( {\frac{{Bt + \frac{{A\,{\rm{sin}}(\omega t)}}
{\omega } - x_0 }}
{{\sqrt 2 \sigma \sqrt t }}} \right) } \right. \hfill \\
 & \left. { - \exp \left[ \frac{{2x_0 (B\omega t + A{\rm{sin}}(\omega t))}}
{{\sigma ^2 \omega t}}\right] {\rm{erfc}}\left({\frac{{Bt + \frac{{A\,{\rm{sin}}(\omega t)}}
{\omega } + x_0 }}
{{\sqrt 2 \sigma \omega \sqrt t }}} \right)} \right\} \hfill \\
\end{array}
\label{eq:sinsurvival}
\end{equation}
\noindent and, analogous to \cite{bulsara1994}, from
equation~(\ref{eq:g1}) the first passage density can be expressed
as
\begin{equation}
\begin{array}{ll}
  g(t\left| {x_0 } \right.) &= \frac{{x_0 \exp \left\{ - \frac{{\left[ {Bt + \frac{{A{\rm{sin(}}\omega t)}}
{\omega } - x_0 } \right]^2 }}
{{2\sigma ^2 t}}\right\} }}
{{\sqrt {2\pi } \sigma t^{\frac{3}
{2}} }}   \hfill \\
  &+\frac{{A\exp \left\{\frac{{[(x_0  + Bt)\omega  + A{\rm{sin(}}\omega t)]^2 }}
{{2\sigma ^2 \omega ^2 t}}\right\} {\rm{erfc}}\left( {\frac{{Bt + \frac{{A{\rm{Sin(}}\omega t)}}
{\omega } + x_0 }}
{{\sqrt 2 \sigma \sqrt t }}} \right)\left[ {t\,{\rm{cos(}}\omega t) - \frac{1}
{\omega }{\rm{sin(}}t\omega )} \right]}}
{{\sigma ^2 t^2 }} \hfill \\
\end{array}
\label{eq:sinFPT1}
\end{equation}
\noindent The approximated nature of the solution is evidenced by
the fact that, the image source intensity is no longer constant
in time, so that by evaluating the probability current in \(0\) we
obtain
\begin{equation}
\tilde g(t|x_0 ) = \frac{{x_0 }}
{{\sqrt {2\pi } \sigma t^{3/2} }}\exp \left\{ - \frac{{[\omega (B t - x_0 ) + A{\rm{sin}}(\omega t)]^2 }}
{{2\omega ^2 \sigma ^2 t}}\right\},
\label{eq:sinFPT2}
\end{equation}
\noindent which is different from~(\ref{eq:sinFPT1}). In any case, the first
passage time pdf in equation~(\ref{eq:sinFPT1}) is in good
agreement with the numerical simulations in
\cite{bulsara1994,bulsara1995}. Also, when \(A \to 0\) both
the~(\ref{eq:sinFPT1})  and the~(\ref{eq:sinFPT2}) tend to the
first passage time pdf for a simple biased Bm.

\begin{figure}
\noindent\includegraphics[width=35pc]{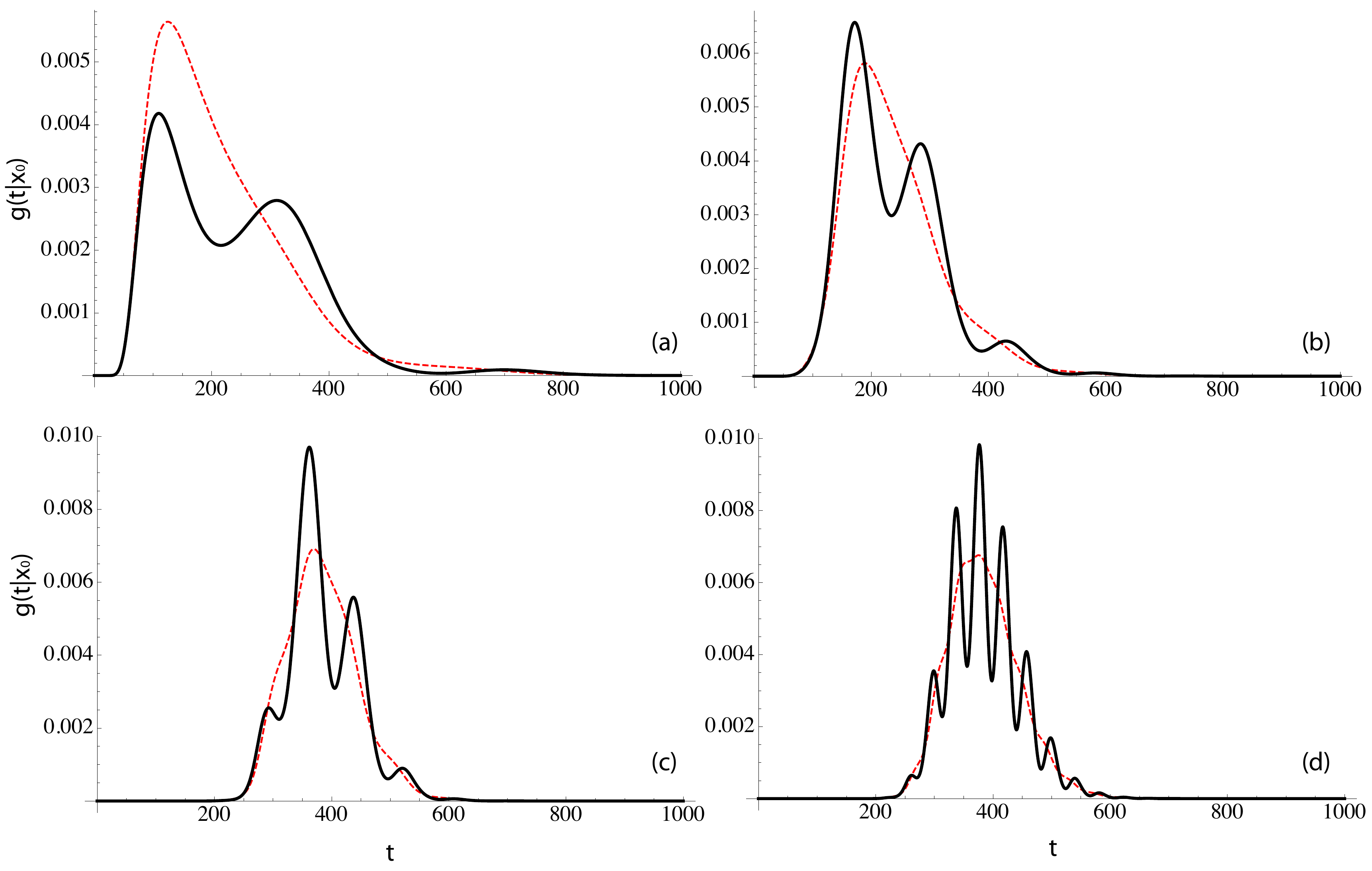}
\caption{First passage densities \(g(t|x_0 )\) (bold black line, Eq.~\ref{eq:sinFPT1}), and \(\tilde g(t|x_0 ) \) (red dotted line, Eq.~\ref{eq:sinFPT2}), respectively for (a) \(\mu=0.065\), \(\sigma=0.5\), \(x_0=25\), \(A=0.032\) and \(\omega=0.016\); (b) \(\mu=0.065\), \(\sigma=0.35\), \(x_0=15.5\), \(A=0.025\) and \(\omega=0.04\); (c) \(\mu=0.065\), \(\sigma=0.2\), \(x_0=25\), \(A=0.03\) and \(\omega=0.07\), and (d) \(\mu=0.065\), \(\sigma=0.2\), \(x_0=25\), \(A=0.03\) and \(\omega=0.15\). The discrepancy between $\tilde{g}(t|x_0)$ and $g(t|x_0)$ clearly signifies the failure of the method of images for problems with time-dependent P{\'e}clet numbers.}
 \label{FStochasticResonance}
\end{figure}

As highlighted in Figure~(\ref{FStochasticResonance}), when the
magnitude of $\mu(t)$ becomes significant, the two pdfs diverge
due to the losses of probability density at the barrier
(Eq.~(\ref{eq:sinFPT2})). For this reason, the method of
images cannot be considered a general approach to solving problems
described by Eq. (\ref{eq:fpe}) with a time-dependent
P{\'e}clet number.


\section{A Case Study: Snowmelt Dynamics} 
\label{sec:snowmelt} Snowmelt represents one of the paramount
sources of freshwater for many regions of the world, and is sensitive to both temperature and
precipitation fluctuations \citep{,Barnett2004,barnett2005,Barnett2009,perona2001, pepin2008,you2010}.
Snow dynamics is characterized by an accumulation phase during
which snow water equivalent (i.e. the amount of liquid water potentially available by totally and instantaneously melting the entire snowpack)
increases until a seasonal maximum $h_0$ is
reached, followed by a depletion phase in which the snow mantel
gradually decays (and releases the stored water content) due to the increasing air temperature. 
Such a dynamics is complex and its general description requires numerous physical
parameters that are rarely measured or available. In this
section, we focus on a stochastic model describing the total water equivalent from both snow and 
rainfall during the melting season,  as forced/fed by both precipitation 
(moving from the solid to the liquid precipitation regime) and increasing air temperature. 

Due to the simplified nature of our stochastic model, we will consider the total potential water availability (in terms of water equivalent) as the key variable, 
thus neglecting any further effects connected with snow percolation and metamorphism \citep{Dewalle2008}.
Snowfalls are here assumed to become more sporadic progressing into the warm season and the predominant 
controls over fresh water availability during the melting period are increasing air temperature and liquid precipitation.
Accordingly, the melting phase is described by a power-law time dependent drift
directed towards the total depletion of the snow mantle and by a
power-law diffusion whose positive and negative excursions
represent respectively precipitation events and pure melting
periods. The melting process is often described by a linear function of time by using the so called ``degree-day'' approach with time-varying melting-rate coefficients \citep{Dewalle2008}.
Considering that temperature varies seasonally and increases during the melting season, a power-law form for drift and diffusion during the spring season, still represents a parametrically-parsimonious and effective approximation of the basic driver of the process. 

\begin{figure}
\noindent\includegraphics[width=35pc]{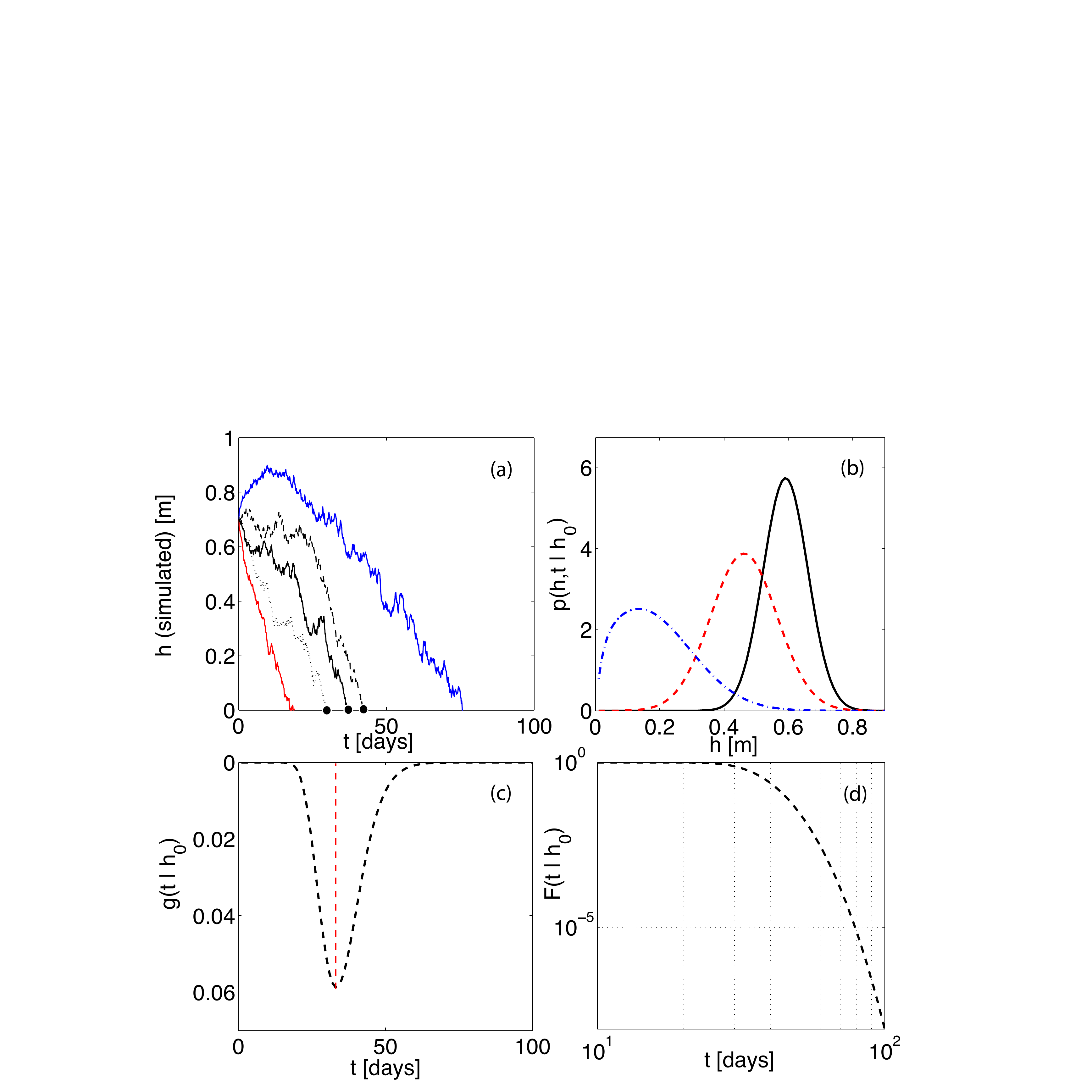}
\caption{Sample trajectories of specific water equivalent from elevated regions during the melting season  (a), and analytical results for the coupled stochastic melting-precipitation process of equation (\ref{eq:langevin}) (panels (b) to (d)). Numerical results were obtained by simulating Eq. (\ref{eq:langevin}) by means of an Euler algorithm with step $10^{-2}$ days. Panel (a) shows few sample trajectories of the process together with the curve of maximum values (upper curve) and minimum values (lower curve) over an ensemble of $10000$ simulations, for $\alpha=0.25$ and $k=0.24$ ${\rm mm^2/days^\alpha}$. Analytical results for the conditional probability $p(h,t|h_0)$ at different instants, the first passage time density $g(t|h_0)$, and the survival function $F(t|h_0)$, are also shown in panels (b) to (d).}
 \label{FSnow}
\end{figure}

Under these assumptions, the dynamics of the total water equivalent depth for unit of area  $h$ -- i.e. the amount of fresh water potentially available from both snow accumulation and rainfall \citep{Bras1990} -- at
a given point in space can be can be reasonably described by the Langevin
equation
\begin{equation}
dh=  - qkt^\alpha dt   + \sqrt {2kt^\alpha} dW(t)
\label{eq:langevin}
\end{equation}

\noindent where $k$ (with dimension $L^2/T^{\alpha+1}$) represents the
accumulation/ablation rate. Note that here $h$ includes both the rainfall and snowmelt contributions.
Also, we hypothesize that both the drift and the diffusion scale with the same exponent $\alpha$.
This is a reasonable assumption given that variability of 
the process is expected to increase proceeding into the warm season.
The initial condition is given by the snow water equivalent ($SWE$)
$h_0$, accumulated during the cold season. The
survival probability $F(t|h_0)$ for a given initial $SWE$
and the first passage time density $g(t|h_0)$ can be respectively
calculated from (\ref {eq:Spowerlaw}) and (\ref{eq:gpowerlaw}).
Figure~\ref{FSnow} shows few sample trajectories of the process
(panel (a)) obtained by the numerical simulation of Eq.
(\ref{eq:langevin}) by means of a forward Euler algorithm with a
time step of $10^{-2}$ days. The conditional probability $p(h,t|h_0)$ at different
instants, the first passage time density $g(t|h_0)$, and the survival function $F(t|h_0)$, for
the case $\alpha=0.25$ and $k=0.24$ ${\rm mm^2/days^\alpha}$ are also shown in panels (b) to (d). 
Here, we calibrated the parameters to obtain the mode of the first
passage time at about 40 days after reaching the maximum $SWE$ of
the season $h_0$.
The first passage time statistics presented offer important
clues about the timing between melting and summer fresh-water
availability under different climatic scenarios (consider for
example the FPT pdf in Figure ~\ref{FSnow}(c)).


\section{Conclusions}
\label{sec:conclusions} The first passage time properties of
Brownian motion with purely time dependent drift and diffusion
coefficients subjected to an absorbing barrier were investigated.
These processes can be used to mimic a variety of environmental
and geophysical phenomena, representing the availability of a resource and its
dynamics in time (e.g. the ablation phase of a snow mass
accumulated during the winter period and forced by temperature and
precipitation). Survival functions and pdfÕs for the first
passage times at the barrier were derived for power-law and
periodic forcing time-dependent drift and diffusion terms for the
associated Fokker Planck equation using the method of images. The
general properties and limitations of this method were also
reviewed, with reference to previous results obtained in the field
of neural sciences and stochastic resonance. Particularly, we
discussed how the applicability of the method of images to a Bm with
time-dependent drift and diffusion is limited to the case of a
process with constant P{\'e}clet number, i.e. with a time-independent
ratio of drift and diffusion.

Where the time dependence
is of the power-law type, the derived first passage time density
and survival functions share many analogies with the statistics of
inter arrival times between intermittent events when the
considered system displays sporadic randomness. In the case of a
periodic time-dependence, first passage time statistics appear to
be modulated by the frequency of the forcing. The periodic forcing
case has been also used to show the approximate nature of
solutions obtained by the method of images, when time-dependent
drift and diffusion terms are not linearly related. We finally
show how a Bm with power-law decaying drift and diffusion can be
used to describe the warm season dynamics of the total water equivalent in
mountainous regions.

\section{Acknowledgments}
This study was supported, in part, by the National Science
Foundation (NSF-EAR 0628342, NSF-EAR 0635787 and NSF-ATM-0724088),
and the Bi-national Agricultural Research and Development (BARD)
Fund (IS-3861-96). We wish to thank Adi Bulsara for the helpful suggestions.
We also thank Demetris Koutsoyiannis and the other three 
anonymous reviewers for their helpful suggestions.

\bibliographystyle{model5-names}
\bibliography{PHYSA_SDE_BIBLIO}
\newpage







\end{document}